\documentclass[aps,prl,twocolumn,superscriptaddress]{revtex4-1}

\usepackage[text={15.5cm,23cm},centering]{geometry}

\usepackage{mathptmx}
\usepackage{graphicx}
\usepackage{subfigure}
\usepackage{color}

\usepackage{xcolor}
\usepackage{hyperref}
\usepackage{amsmath}

\begin{document}
\title{Emergent quantum matter in graphene nanoribbons}

\begin{abstract}
In this book chapter, we introduce different schemes to create quantum states of
	matter in engineered graphene nanoribbons. We will focus on the
	emergence of controllable magnetic interactions, topological quantum
	magnets, and the interplay of magnetism and superconductivity. We
	comment on the experimental signatures of those states stemming from
	their electronic and spin excitations, that can be observed with atomic
	resolution using scanning probe techniques.
\end{abstract}

\author{J. L. Lado}
\affiliation{Institute for Theoretical Physics, ETH Zurich, 8093 Zurich, Switzerland }
\affiliation{Department of Applied Physics, Aalto University, Espoo, Finland}
\author{R. Ortiz}
\affiliation{Departamento de F\'isica Aplicada,  Universidad de Alicante, 03690 Spain}  
\author{J. Fern\'andez-Rossier}
\affiliation{QuantaLab, International Iberian Nanotechnology Laboratory,
Avenida Mestre Jos\'e Veiga, 4715-330 Braga, Portugal}

\maketitle

\section{Introduction}

In this chapter we provide a perspective on the potential of graphene
nanoribbons as a platform to host a variety of non-trivial emergent electronic
states, such as  topological phases,  quantum spin liquids,
broken symmetry magnetic states and Yu-Shiba-Rusinov excitations.
 This potential arises from the capability to nanoengineer the electronic
 properties of nanoribbons using several different resources:
 \begin{itemize}
 \item Geometrical control. Graphene ribbons with different shapes,
	 orientation, widths, can be synthesized\cite{wang16}. This 
		 gives rise to different electronic properties, including the
		 emergence of localized edge and interface states that can host
		 unpaired spin electrons
 \item  Tuning the electron density.  Using either gating and chemical doping it is possible to control the density of electrons.
 \item  The electronic properties of GNR can be affected by several types of proximity effect: spin-orbit, superconducting and magnetic.   Therefore, they provide an unique platform to explore the interplay between local magnetic order and superconductivity.
 \end{itemize}
 
 In addition,  scanning probe spectroscopies permit to probe both the
 structural properties of GNR as well as their electronic
 properties\cite{tao11,ruffieux12,ruffieux16,Jacobse2017,Schulz2017} and 
 spin excitations\cite{li19},
 with atomic scale resolution, and constitute great tool to probe the emergent
 electronic phases.\cite{2019arXiv190503328Y}

The most compelling argument to expect non-trivial correlated phases in GNR
comes from experiments. Non-trivial phases, including Mott-Hubbard insulating
phases\cite{cao18a} and non-trivial superconductivity\cite{cao18b} have been
observed in  twisted graphene bilayers.  Whereas the precise origin of the
superconductivity is not understood, there is a consensus on the crucial role
played by an array of localized states that form very narrow bands
  at the Dirac energy.  Given that the chemical properties of monolayer GNR are
  almost identical to those of graphene bilayer,  it is our contention that the
  localized zero modes of the GNR can also result in non-trivial correlated
  phases.  
The scope of the chapter is to unveil
some physical mechanisms that can promote non-trivial electronic phases, rather than the technical aspects of how to model them.

We focus on three  classes of non-trivial electronic behavior. First, we
revisit the thoroughly studied problem of magnetic order in the zigzag edges.
We address the prominent role of spin fluctuations in this low dimensional
system.  Second, we discuss how to
realize spin chain Hamiltonians in graphene ribbons engineered to host  an
ordered array of localized zero mode states.  Third, we address the interplay
between emergent local moments and superconducting  proximity effect.

\section{Modeling GNR}
\subsection{Geometries}
In this chapter we focus on GNR with atomically precise edges along the crystallographic axis of graphene.  These are the so  called armchair and zigzag edges.  We consider both finite (0D) and infinite (1D) ribbons. We also consider ribbons with a periodic modulation of their width,  that are known to host interfacial topological zero modes\cite{Grning2018,cao17,ortiz18,rizzo18}.  Several other geometries, such as chiral ribbons with sufficiently long zigzag patches, as well as chevron type ribbons with zigzag edges, can result in the formation of local moments and non-trivial spin physics. 

\begin{figure*}[]
\begin{center}
\includegraphics[width=0.8\textwidth]{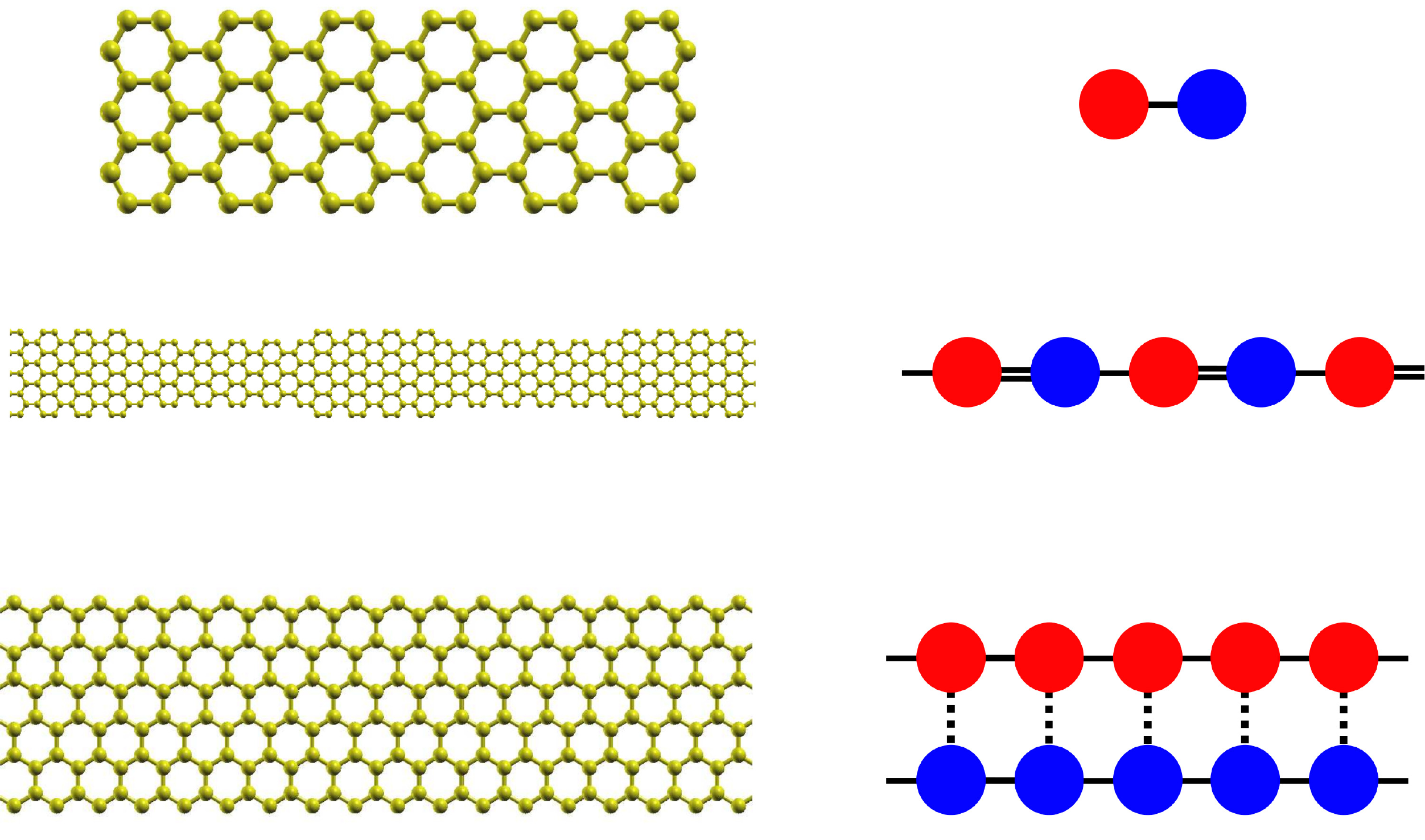}
\end{center}
\label{fig0}
\caption{Left column different GNR considered in this chapter. Top: a finite size graphene ribbon.  Middle: 1D graphene ribbon that alternates to sections with different width and armchair edge.  Bottom: a 1D ribbon with zigzag edges.   Right column: the equivalent lattice spin model. }
\end{figure*}

\subsection{Single particle terms}
Here we adopt the standard model \cite{neto09,katsnelson12} to describe GNR, namely, a single orbital tight-binding model with first neighbor hopping $t$.  The single orbital is the $p_z$ atomic orbital of carbon, that is decoupled from the rest in planar structures. 
Unless otherwise stated, we assume that edge carbon atoms are passivated with hydrogen, so that there are no dangling bonds at the surface.  This model gives a fair description of the states in a few eV window around the Fermi energy for planar
carbon based structures, going from zero dimensional molecules, to planar graphene. 
 It is standard\cite{neto09}
 to take  $t=-2.7 eV$, which provides a good slope for the Dirac
 cones in planar graphene. 
The first neighbor hopping Hamiltonian reads:
\begin{equation}
{\cal H}_0= -t \sum_{i,i',\sigma} c^{\dagger}_{i\sigma}c_{i',\sigma}
\label{h0}
\end{equation}
where $i'$ stand for the first neighbors of $i$.  This is the dominant term in the Hamiltonian and  it  accounts for the Dirac cones in graphene \cite{neto09,katsnelson12}, the existence of localized states in the zigzag edges\cite{nakada96},  and the gapped bands in  armchair  GNR\cite{nakada96}.  

In addition, we consider the effect of the several spin-dependent terms in the Hamiltonian. First, 
the Zeeman term, given by
\begin{equation}
{\cal H}_Z =\frac{1}{2}g\mu_B \vec B \cdot\sum_{i,\sigma,\sigma'} \vec \sigma_{\sigma,\sigma'} 
 c^{\dagger}_{i\sigma}c_{i,\sigma'}
\label{Zeeman}
\end{equation}
where $g\simeq 2$ and $\vec \sigma_{\sigma,\sigma'}$ are the Pauli matrices. 

The  intrinsic spin-orbit coupling, proposed by Kane and Mele, is described by
\cite{kane05}: 
\begin{equation}
H_{KM} = 
\sum_{i,i'' \sigma}
it_{KM} \sigma\nu_{i,i''}c^{\dagger}_{i\sigma}c_{i''\sigma}
\label{Hamil}
\end{equation}
where $i''$ stands for the second neighbors of $i$, summation, $\sigma=\pm1$
are the spin projections (along the axis perpendicular to the crystal plane)
and $\nu_{i,i''}=+(-)1$ for clockwise (anticlockwise) second neighbor hopping. 
  When added to the  hopping Hamiltonian (\ref{h0}) , the  Kane-Mele term opens a
  topologically non-trivial  band-gap $\Delta_{SOC}= 6\sqrt{3}t_{KM}$  at the
  Dirac points.     The non-trivial nature of the gap implies the emergence of
  spin-locked chiral edge states\cite{kane05}.   Because of the small magnitude
  of   $\Delta_{SOC}$ in graphene, the observation of this gap is very
  challenging\cite{PhysRevLett.122.046403} and the
  localization length of the edge states is very large.
  Therefore, this term has a minor influence in the properties of graphene
  ribbons.  However, this type of term could be enhanced by proximity
  effect\cite{kou15}.

A second type of spin obit effect can arise when 
  mirror symmetry is broken,  due to the application of an external off-plane
  electric field, or due to interaction with the substrate. This is the so
  called Rashba spin-orbit term \cite{kane05,min06}:
\begin{equation}
{\cal H}_R = it_R \sum_{i,j,s,s'} \vec E \cdot \left ( \vec r_{ij} \times \vec \sigma \right )_{s,s'}
c_{is}^\dagger c_{js'}
\label{Rashba}
\end{equation}
where $\vec r_{ij}$ is unit vector along the bond  between the carbon sites $i$ and $j$, 
$\vec \sigma$ are the spin Pauli matrices and $\vec{E}$ is a vector related to inversion symmetry breaking of the graphene lattice, such as an off-plane electric field\cite{min06,ortiz18}. The Rashba spin orbit coupling does not commute with $S_z$ and promotes mixing between the two spin channels. 

\subsection{Coulomb interaction}
In this chapter we consider the effect of electron-electron Coulomb interactions within the Hubbard approximation:
\begin{equation}
{\cal H}_{U} = U\displaystyle\sum_{i}n_{i\uparrow}n_{i\downarrow}
\end{equation}
where $U$ stands for the Coulomb penalty for having 2 electrons in the same
$\pi$ orbital in a single carbon atom. 
The value of $U$ may depend on additional screening effects, 
including the substrate. In addition, the right value of
$U$ might depend on whether or not we include a next-neighbor Coulomb repulsion
in the Hamiltonian\cite{PhysRevLett.106.236805}.
Here we adopt $U$ as a variable parameter, that
takes values in the range of $U=|t|$.

The Hubbard model can only be solved   exactly in very specific geometries,  such as 
the  monostrand one dimensional chain. Thus, very often\cite{fujita96,fernandez07,fernandez08,munoz09,lado14,lado14b,lopez17} the model is treated at the mean field approximation, where the exact Hamiltonian is replaced by an effective Hamiltonian
\begin{equation}
{\cal H}_{U,MF} = {\cal H}_{\rm Hartree} + {\cal H}_{\rm Fock}
\label{MF}
\end{equation}
where 
\begin{equation}
{\cal H}_{\rm Hartree} = 
U \left( n_{i,\uparrow}\langle n_{i,\downarrow} \rangle
+
n_{i,\downarrow} \langle n_{i,\uparrow} \rangle
\right)
\label{Hartree}
\end{equation}

\begin{equation}
{\cal H}_{\rm Fock } = 
-U\left(
c_{i,\downarrow}^\dagger c_{i,\uparrow}
\langle
c_{i,\uparrow}^\dagger c_{i,\downarrow}
\rangle
+
c_{i,\uparrow}^\dagger c_{i,\downarrow}
\langle
c_{i,\downarrow}^\dagger c_{i,\uparrow}
\rangle
\right)
\label{Fock}
\end{equation}
so that electrons interact with an external field that is
self-consistently calculated.  Most often \cite{fujita96,fernandez07,fernandez08,jung09,munoz09,yazyev10,soriano12} an additional approximation has been used, that assumes  a collinear magnetization so that the Fock term vanishes.
 For small nanographenes,   such as triangular and hexagonal islands with
 zigzag edges\cite{fernandez07},  the results of  collinear mean-field calculations of
 the Hubbard model  are very similar to those obtained using   density functional theory calculations that
 include long-range Coulomb interactions  and include several atomic orbitals per carbon atom.
  The
 same statement holds true for infinitely long graphene ribbons with zigzag edges: 
both  mean field calculations Hubbard model
calculations\cite{fernandez08,fujita96} and DFT based calculations\cite{son06}
predict ferromagnetic order at the edges and antiferromagnetic inter-edge
coupling at half filling.

The study of non-collinear magnetization has permitted to study the canted spin phases in graphene quantum Hall bars \cite{lado14b} as well as the existence of in-gap topological fractional excitations at the domain walls of graphene zigzag ribbons \cite{lopez17}.

 \subsection{Proximity terms}
 The effective Hamiltonian for electrons in graphene can be
 modified due to the interaction with the substrate.  The most frequently
 considered types of proximity terms are a sublattice symmetry-breaking on-site
 potential, that opens up a gap\cite{soriano12,giovannetti07}, a ferromagnetic
 spin proximity effect, that splits the bands
 \cite{qiao10,yang13,hallal17,cardoso18} and a superconducting proximity
 effect, that adds a pairing term to the Hamiltonian, and opens up a
 superconducting gap to graphene whenever the Fermi energy lies on a band.

  The on-site potential   can be written down as:  
\begin{equation}
H_J =  \sum_{i } W(i) c_{i\sigma }^\dagger c_{i\sigma} 
\end{equation}
Whenever $W(i)$ is different for $A$ and $B$ sublattice, this term can open up
a gap in graphene.  When the sign of this gap is modulated across graphene,
kink states can emerge\cite{PhysRevD.13.3398}.  
 
 The spin proximity effect can be written down as:  
\begin{equation}
H_J = \frac{1}{2} \sum_{i }  \vec{J}(i)\cdot\vec{\sigma}_{\sigma,\sigma'} c_{i\sigma }^\dagger c_{i\sigma'} 
\end{equation}
where $\vec{J}(i)$ is the exchange field that is proportional to the
magnetization field of the proximity layer.    In the simplest scenario,  this
is taken as a collinear and constant field, so that the spin proximity effect
leads to a spin splitting of the bands that, in conjunction with Rashba spin
orbit coupling, can induce a quantized anomalous Hall phase\cite{qiao10}.
Spin proximity with non-collinear or even non-coplanar substrates, such as
skyrmions, can also result in a quantized anomalous Hall phase, without the
need of spin orbit coupling\cite{lado15b}.  The typical magnitude for the
exchange splitting, as obtained from DFT
calculations\cite{yang13,hallal17,cardoso18},  is in the range of a few tens of
meV at most.  Experimentally,  a report of splitting induced by spin
proximity effect observed in graphene is much smaller than that, in the range
of a fraction of meV \cite{leutenantsmeyer16}. 

The  superconducting proximity  effect is introduced as an
effective conventional s-wave pairing term:
\begin{equation}
H_{SC} =  \Delta \sum_{i} \left [ c_{i,\uparrow}c_{i,\downarrow} +
c^{\dagger}_{i,\downarrow} c^{\dagger}_{i,\uparrow} \right ]
\end{equation}
This term  has to be treated using the so called Bogoliubov de Gennes  (BdG)  Hamiltonian\cite{beenakker06}. 
 
 \begin{figure*}[]
\begin{center}
\includegraphics[width=\textwidth]{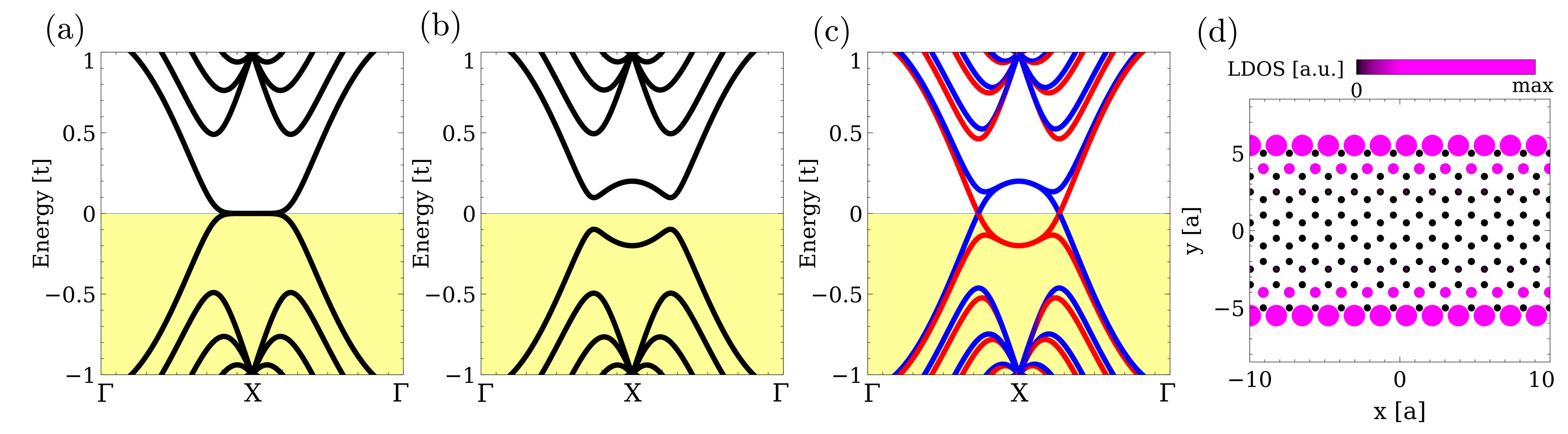}
\end{center}
	 \caption{(a) Band structure of a zigzag nanoribbon with a 16 atoms
	 width. Band structure with local exchange fields in the
	 upper and lower edges, aligned antiferromagnetically (b) and
	 ferromagnetically (c). (d) Spatial
	 distribution of the flat band of (a), that become
	 magnetized in panels (b,c). }
\label{figzz1}
\end{figure*}

\section{Emergent phases and Zero modes}

\subsection{Single particle theory of Zero modes}

We consider graphene ribbons not too far from their charge neutrality point. Therefore, their Fermi energy lies close to the Dirac point.  Because of quantum confinement,  extended states of graphene ribbons are gapped. This leads to semiconducting or insulating ribbons that are not expected to host non-trivial  electronic phases.   The way out of this situation comes from the existence of  zero modes.   In pristine GNR,  zero modes  arise in the following instances:
\begin{enumerate}
\item At sufficiently long zigzag edges\cite{nakada96}. As we discuss below, there is one zero mode for every three carbon atoms in a zigzag edge. 
\item At interfaces between gapped armchair ribbons with different symmetry protected topological indexes $Z_2$, defined below, as proposed by Cao {\em et al.} \cite{cao17}
\end{enumerate}
In addition,  zero modes also appear when graphene is functionalized
with atomic hydrogen\cite{yazyev07,palacios08,gonzalez16,garcia17} or any other
sp$^3$ functionalization\cite{santos12}. 

\begin{figure*}[t!]
\begin{center}
\includegraphics[width=0.9\textwidth]{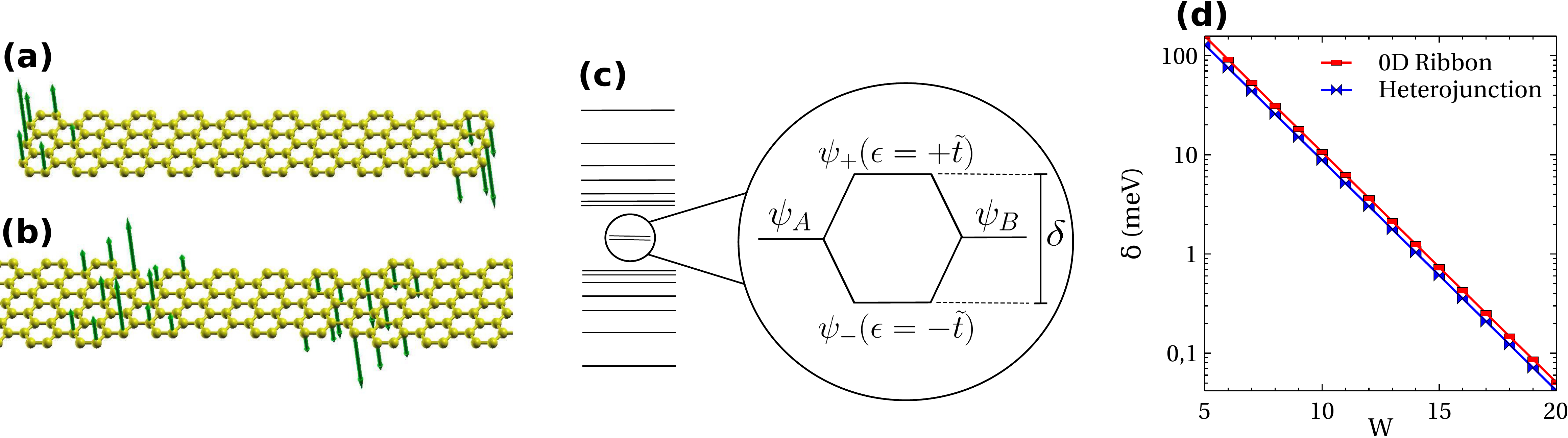}
\end{center}
\caption{Left:  GNR with short zigzag edges (top) and aGNR with two widths.
	 Both structures have in-gap states that host local moments when
	 Coulomb interactions are included. The vector maps reflect the
	 magnetization obtained in a mean field calculation with the Hubbard
	 model.  
Right:
Scheme of the single-particle $U=0$ energy spectrum showing two almost
	 degenerate in-gap states. Their wave functions are the symmetric and
	 antisymmetric combinations of the sublattice polarized zero modes
	 located at the edges/interfaces. The exponential dependence of the
	 hybridization, as measured by the in-gap splitting, as a function of
	 $W$, the length scale that controls the size of the GNR.   }
\label{figzeromode}
\end{figure*}

 There are two complementary ways to understand the emergence of  these zero modes.  The first way   invokes the bipartite character of the honeycomb lattice and the emergence of  at least $N_A-N_B$ zero modes\cite{sutherland86,palacios08}, where  $N_A$ and $N_B$ are  number of sites in the two sublattices that form a structure. In addition, the theorem permits to anticipate the sublattice polarized nature of the zero modes. 
   This first method  permits to predict the emergence of zero modes in 
 sp$^3$ functionalized graphene\cite{palacios08}. There, the $p_z$ orbital
 forms a strong covalent bond with an orbital of the functionalizing species,
 such as the $1s$ orbital of atomic hydrogen.  This takes away both 1 electron
 and 1 orbital from the $p_z$ array.  This can be effectively modeled as a
 tight-binding model with a missing site\cite{palacios08,soriano10}.   The
 sublattice imbalance argument can also be applied right away to the interface
 states between armchair ribbons,\cite{ortiz18}
 shown in figure \ref{figzeromode}.
 In the case of graphene zigzag edges, it can be
 invoked, although in a less rigorous manner.  Locally, zigzag edges have
 sublattice imbalance, but globally, the structures  have $N_A=N_B$.
 
 The existence of zero modes in some GNR can also be related to topological
 arguments. The interface between two media described with different
 topological indexes, $N_1$ and $N_2$, is expected to host at least $N_1-N_2$
 zero modes.  In a 1D crystal with mirror and inversion symmetry, 
 we define the Zak phase of a band $n$ as\cite{cao17,delplace11}:
 \begin{equation}
\gamma_n= i\left(\frac{2\pi}{d}\right)\int_{-\pi/d}^{\pi/d} dk \langle u_{nk}|\frac{\partial u_{nk}}{\partial k}\rangle
 \end{equation}
  where
   $d$ is the unit cell size and $u_{nk}$ is the periodic part of the Bloch wave function for band $n$. 
   In a symmetry protected 1D crystal, the Zak phase is quantized as $0$ or $\pi$ modulo 2$\pi$. This permits to define a topological index:
   \begin{equation}
   (-1)^{Z_2}=e^{i\sum_n\gamma_n}
   \end{equation}
From the bulk-boundary correspondence,  symmetric junctions of armchair ribbon with different Z2 numbers are expected to host localized zero modes at the interfaces. This has been confirmed both with DFT \cite{cao17} and tight-binding calculations\cite{ortiz18}. These junctions happen to have $|N_A-N_B|=1$, so that the  interface zero mode can be understood using the theorem for bipartite lattice.  Using similar arguments\cite{delplace11}, the existence of edge modes in 1D zigzag edges has been related to the Zak phase for the family of 1D states defined in a cut of the 2D Brillouin zone.

\subsection{Infinite ribbons}
We begin our discussion of specific systems with the case of   one dimensional
graphene ribbon with zigzag edges. As shown in figure \ref{figzz1}(a), the
energy bands feature two flat bands at $E=0$. 
These two
bands of zero modes occupy exactly one third of the Brillouin zone. Given that
the unit cell of the ZZ GNR has exactly one carbon site per edge,  this implies
that the ratio  of zero modes per carbon edge atom is $1/3$.   The wave
function of the edge modes is sublattice polarized, and its amplitude
quickly decays
as we move inwards in the GNR.   Other than these zero modes, the
rest of the bands are gapped,  reflecting the confinement of the Dirac
particles in the section of the ribbon. 

The flat bands at $E=0$ give rise to a very large density of states at that
energy.  Given that  $E=0$ is the Fermi energy for half filling,  interactions
are expected to have a strong impact in this system. This was found out more
before the turn of the century, using a mean field approximation for the
Hubbard model in this system\cite{fujita96} and subsequent work, using both the
Hubbard model\cite{fernandez08,munoz09,lado14} and DFT calculations.
In all
instances, the predictions of these symmetry breaking methods are:
\begin{itemize}
\item The  zigzag edges are ferrromagnetic, 
with  magnetic moments in the range of 0.15 $\mu_B$ per carbon atom
\item The inter-edge interaction is antiferromagnetic and decays rapidly as a function of the width
\item The energy bands, 
	show a dispersion of the edge states, driven by the interactions\cite{fernandez08}.  
In the case of parallel (antiparallel)  alignment of the edge
	magnetizations,  the  ZZ is a conductor (insulator). This finding
		prompted proposals for using graphene ribbons as ideal spin
		valves\cite{kim08,munoz09}.
\end{itemize}

The qualitative effect of interactions can be captured
by adding local exchange fields at the upper and lower
zigzag edge, giving rise to results comparable
to the full selfconsistent calculation (Fig. \ref{figzz1}(b,c)). 

In general, mean-field calculations  for any nanographene  with zigzag edges
predicts the existence of magnetic moments  localized at the edges with
ferromagnetic correlations between edges that belong to the same sublattice,
and antiferromagnetic correlations between edges that belong to opposite
sublattices\cite{fernandez07}.  
In the case of infinite 1D ribbons, these calculations have an obvious problem: they predict infinitely long range
order along the edge,  breaking a continuous symmetry in one dimension. This is
incompatible with well established theorems. In one dimension, quantum
fluctuations are known  to destroy this type of long range order. Therefore, we
need to carry out a treatment that models this system without this drawback.
Before doing that, a possible way out would be to include the terms in the
Hamiltonian that break the SU(2) spin symmetry, given that 1D order is possible
at $T=0$ in Ising chains, for instance. It has been shown\cite{lado14} that
intrinsic spin orbit coupling  favors in-plane edge magnetization.  As a
result, the group of symmetry is reduced, but is still a continuous O(2)
symmetry, for which no long range order can exist in 1D.  In addition,  the
value of the magnetic anisotropy scales with the square of the intrinsic
spin-orbit coupling term in the Kane-Mele Hamiltonian, which is in the range of
a few tens of $\mu$eV in graphene\cite{min06}. Therefore, the magnetic
anisotropy driven by the intrinsic spin orbit coupling in graphene is 
negligible. 

The effect of spin wave fluctuations was considered by Yazyev and Katnelson
\cite{yazyev08}. They computed the spin correlation functions along the edges
using a spin ladder model  of an infinite ribbon and found a power law decay,
with temperature dependence spin correlation length. A Quantum Monte Carlo
description for the same  a spin ladder model for zGNR was also carried out
\cite{koop17}.
 Beyond mean field explorations of  edge ferromagnetism in zGNR  has also been addressed  out with fermionic models including  long range Coulomb interactions using both exact diagonalizations in a restricted active space in the reciprocal state \cite{shi17} as well as  
Quantum Monte Carlos simulation\cite{raczkowski17}. Both methods confirm  intra-edge ferromagnetic correlations. 
In any event, it is apparent that  a rigorous quantum theory for the edge magnetism
has to go beyond  broken symmetry solutions in order to include  a proper treatment of quantum fluctuations. 

\subsection{Finite ribbons} 
We now consider two different finite size ribbons (see figure
\ref{figzeromode}).  Both of them have two weakly hybridized zero modes. The
first one is a ribbon  with two long armchair edges and two short  zigzag
edges, shown in figure \ref{figzeromode}, that  hosts just one 
edge state each.  The second structure combines armchair ribbons with
different width and mirror symmetry, such that the interface hosts zero modes.
For these structure, we take
periodic boundary conditions so that there are no free zigzag edges.
In both cases,  the structures have two zero modes inside a quite large gap.
At half filling, the two edge modes host one electron.  We can treat these
systems, including interactions,  by  considering configurations where the
valence state are doubly occupied and the conduction states are empty.
Therefore, we have a problem of 2 electrons in two sites, that can be solved
analytically\cite{ortiz18}. In both structures, we can change the dimensions of
the system,  and thereby $W$,  defined as the  that the distance   between either the edge or the interface that controls  the hybridization of the zero modes.

{\em  The non-interacting spectrum.}
  A scheme of the single-particle spectrum characteristic of these gapped  0D GNR  with two in-gap states
 is shown in figure 
\ref{figzeromode}(b).  The energies and wave-functions of the in-gap states are
denoted by
$\epsilon_\pm$ and $\psi_\pm$ respectively.
  It is always possible\cite{ortiz18} to write  down the  wave function  of a couple of conjugate states, with single-particle energy $E$ and $-E$,  in terms of the same sublattice polarized states $\psi_A$ and $\psi_B$.  Therefore, we write 
  \begin{eqnarray}
\psi_A(i) \equiv \frac{1}{\sqrt{2}} \left( \psi_+(i) + \psi_-(i)\right) \nonumber\\
\psi_B(i)\equiv  \frac{1}{\sqrt{2}} \left( \psi_+(i) - \psi_-(i)\right) 
\label{zeromodes}
\end{eqnarray} 
 In the case of the in-gap states, the resulting  $\psi_A$ and $\psi_B$
 are {\em spatially separated}. This 
 accounts, in part for the fact that the energy splitting of the zero modes, defined as:
 \begin{equation}
 \delta =2 \langle \psi_A|{\cal H}_0|\psi_B\rangle \equiv  2 \tilde{t}
 \label{zeromodes0}
 \end{equation}
is  small.  In figure \ref{figzeromode}(c) we plot $\delta$ for both the
rectangular  and the heterojunction nanographene, both with two in-gap states.
It is apparent  and well known\cite{nakada96} that this quantity decays
exponentially with $W$.    In the limit where $W$ is very large (see
figure \ref{figzeromode}(c)), $\delta$ vanishes, and the energy of the in-gap
states goes to $E=0$, showing that these sublattice polarized states are
zero modes.\cite{nakada96}

\subsection{$U\neq 0$}
We now consider the effect of interactions and show how it leads to the formation of local moments at the location of the
hybridized zero modes.\cite{golor13,ortiz18}.
The two energy scales that govern the low energy behavior for the two electrons in the two in-gap levels are  $\delta$ and the
energy overhead associated to doubly occupy the sublattice polarized states:
  \begin{equation}
\tilde{U}=U \sum_i |\psi_A(i)|^4 = U\sum_i |\psi_B(i)|^4= U\eta
\label{U}
\end{equation}
The addition energy is thus given by the product of the atomic Hubbard $U$ and $\eta$, is the inverse participation ratio of the zero mode states.  Our numerical calculations for the two structures of figure (\ref{zeromodes})  yield $\eta=0.11$ for the zigzag edge states  and $\eta=0.035$ for the interface states.  We found that, as opposed to the case of $\delta$,  $\eta$ has a very weak dependence of $W$.  We take $U=|t|=2.7 eV$. Therefore, the effective Hubbard interaction $\tilde{U}$ is in the range of 270 and 94 meV, for edge and interface states, respectively.

\subsubsection{Mean field Treatment}

We discuss qualitatively the results of a mean field approximation for the Hubbard model for the two nanographenes of figure(\ref{zeromodes}).   The results are obtained using the collinear 
  mean field treatment (  eqs. (\ref{MF},\ref{Hartree}).    
  For all structures for which   $\tilde{U} \gg \delta$ we  found   broken symmetry solutions with a finite local
magnetization, $M(i)=\langle S_z(i)\rangle$  that are mostly located in the
region where  either $\psi_A$ or  $\psi_B$ are non-zero.  The results of the magnetization field 
are shown in the left panels of figure (\ref{zeromodes}).  The net magnetization per zero mode is close to $S=1/2$. 
  
\begin{figure}[t!]
\begin{center}
\includegraphics[width=0.48\textwidth]{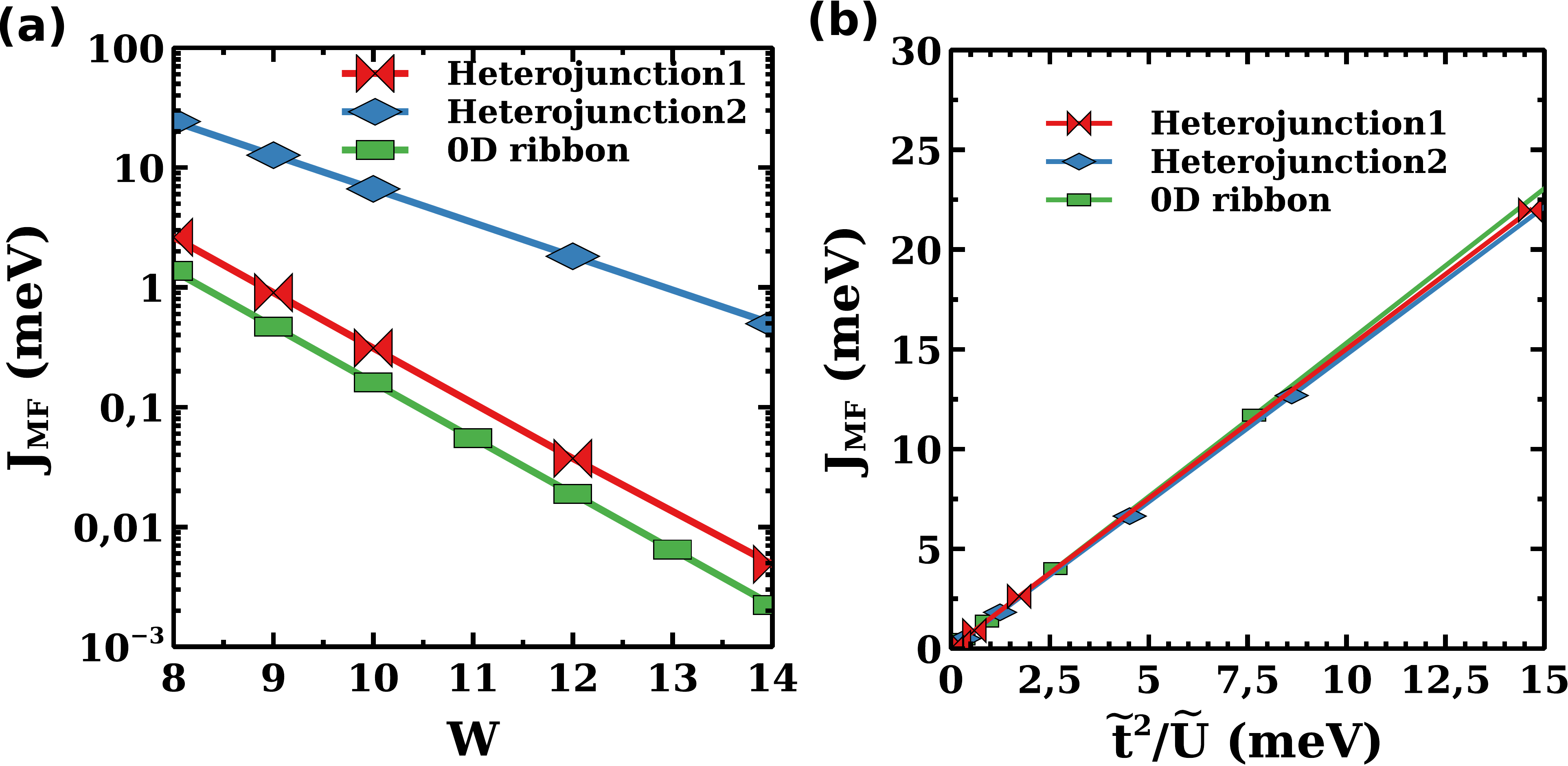}
\end{center}
\caption{Exchange couplings for 0D ribbons and heterojunctions as a function of
	lateral dimension $W$ and as function of
	$\frac{\tilde{t}^2}{\tilde{U}}$. In the case of the heterojunctions, we
	compute independently the coupling mediated by either  the wide  or the
	narrow GNR, by adequate choice of the unit cell dimensions}
	\label{fig:Jribbon}
\end{figure}

Using the mean field approach, we can  study the   exchange energy as the  difference between FM and AF solutions $  J_{MF}= E_{FM}-E_{AF}$
 as a function of  $W$, for both types of structures.  The FM and AF solutions  are obtained by suitably forcing the self-consistent iterative procedure to solve the mean field Hamiltonian. 
   We show in 
figure \ref{fig:Jribbon} that $J_{MF}$ can be as large as 40 meV
can be made small by increasing the distance $W$
between the zero modes.
 Importantly,  as  we show in figure  \ref{fig:Jribbon}(b),  we find that, both
 for
ribbons and heterojunctions, exchange energy scales as
 \begin{equation}
J_{MF}  \propto \frac{\tilde{t}^2}{\tilde{U}}
  \label{KE1} \end{equation}
    This scaling provides a strong indication that the mechanism  of antiferromagnetic interaction is kinetic exchange\cite{anderson59,moriya60}, that arises naturally for half-filled Hubbard dimers. 
The fact that local moments are hosted mostly by the in-gap states permits to build a restricted model where only the in-gap states are considered.  This is the topic of the next paragraph.

\subsubsection{Effective Hubbard dimer}

In order to go beyond the mean
field picture and to be able to  describe  local moments in these nanographenes
with a full quantum theory without  breaking symmetry,  we  
  restrict the Hilbert space  to the configurations of 2 electrons in the two  zero modes.  To do so,   
we represent the Hubbard interaction in the one body basis defined by the states $\psi_A$ and $\psi_B$. The Hamiltonian so obtained is a two site Hubbard model with renormalized hopping and on-site energy\cite{golor13,ortiz18}: 
\begin{equation}
{\cal H}_{\rm eff} = \tilde{t} \sum_{\sigma}\left( a^{\dagger}_{\sigma} b_{\sigma} +  b^{\dagger}_{\sigma} a_{\sigma} \right) 
+  \tilde{U} \left( n_{A\uparrow}  n_{A\downarrow}  + n_{B\uparrow}  n_{B\downarrow}   \right)
\label{H2}
\end{equation}
where $a^{\dagger}_{\sigma}=\sum_i \psi_A(i) c^{\dagger}_{i\sigma}$ and $b^{\dagger}_{\sigma}=\sum_i \psi_B(i) c^{\dagger}_{i\sigma}$ are the  
operators that create an electron in the zero modes $\psi_A$ and $\psi_B$  with spin $\sigma$, respectively. In turn,  $n_{A,\sigma}= a^{\dagger}_{\sigma} a_{\sigma}$ is the number operator for the $\psi_A$ state with spin $\sigma$. 
 
  Hamiltonian (\ref{H2}) is a two-site Hubbard model, where the sites
correspond to the zero mode states $\psi_{A,B}$, shown in figure
\ref{figzeromode}(b, c, d, e).   For the relevant case of 2 electrons,  the
dimension of the Hilbert space is 6 and the ground state is always a singlet, as inferred both from analytical solution 
\cite{jafari08} or by a straight-forward numerical
diagonalization\cite{ortiz18}.    

\begin{figure}[]
\begin{center}
\includegraphics[width=0.5\textwidth]{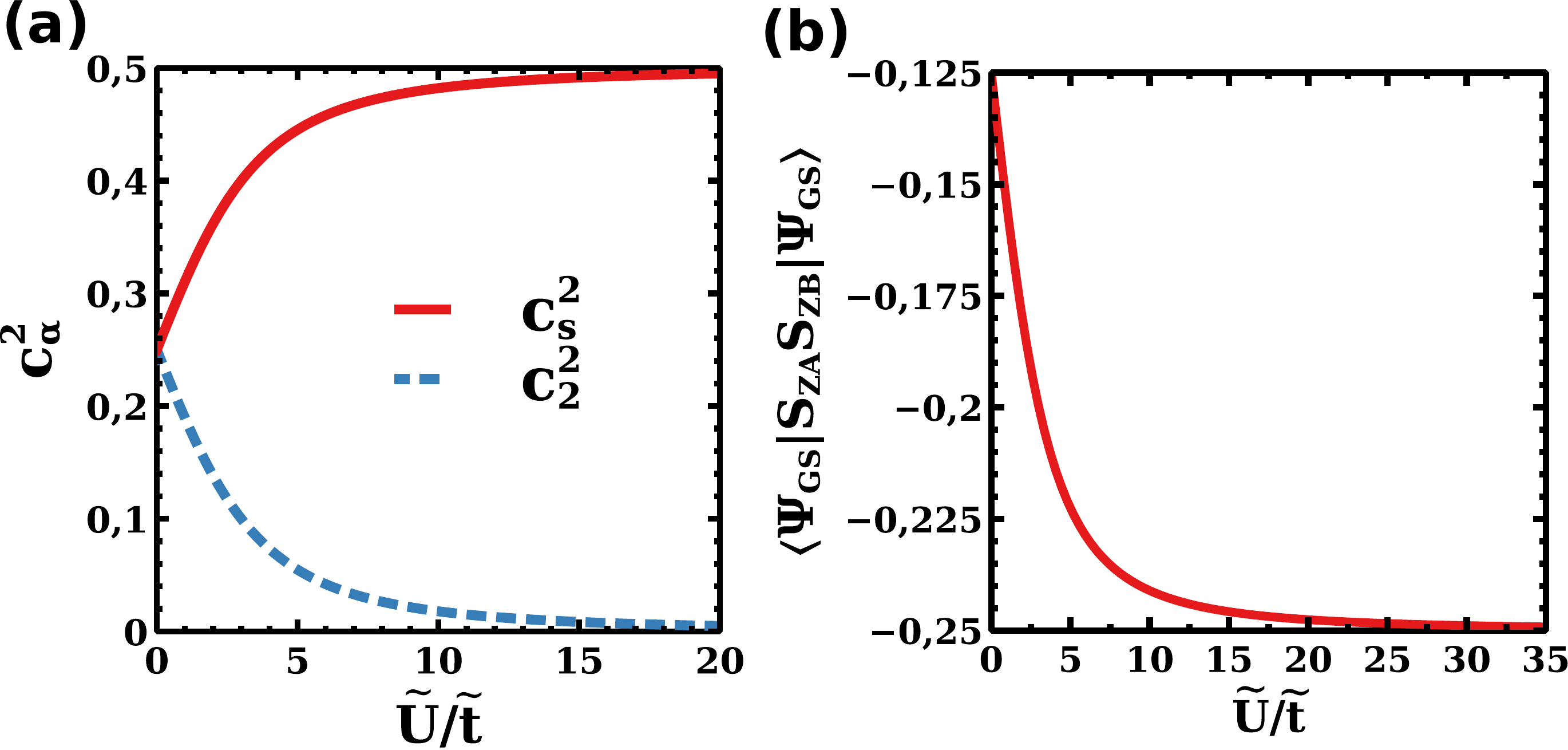}
\end{center}
\label{dimer}
\caption{Left: Evolution of the double occupancy and single occupancy weights in the ground state of the Hubbard dimer at half filling as a function of $U/t$. Right: evolution of the spin correlation function.  }
\end{figure}

The exact solution permits to set the language to discuss the emergence of local moments in these structures. For this matter, we can write the wave function of the ground state as:
\begin{equation}
|\Psi_{GS}\rangle= c_{2} \left(|2,0\rangle +   |0,2\rangle \right)+ c_S\left( |\uparrow,\downarrow>-|\downarrow,\uparrow\rangle\right)
\end{equation}
where $|2,0\rangle$ describes a state with 2 electrons in one site of the Hubbard dimer, and none on the other, whereas $|\sigma_1,\sigma_2\rangle$ describes states with the one electron per site, with spins $\sigma_1,\sigma_2$.
In figure we show hot, for $U=0$, we have $c_2=c_S=\frac{1}{2}$ so that   double occupancy is as likely as  individual occupancy.  As $U$ is ramped up, the $c_2$ coefficient is depleted and the $c_S$ coefficient is enhanced, as shown in figure \ref{dimer}.

In order to characterize the magnetic behavior of the dimer, we define the 
 spin operators:
\begin{eqnarray}
S_z(a)\equiv \frac{1}{2}\left(a^{\dagger}_{\uparrow}a_{\uparrow}-a^{\dagger}_{\downarrow}a_{\downarrow}\right)\nonumber \\
S_z(b)\equiv \frac{1}{2}\left(b^{\dagger}_{\uparrow}b_{\uparrow}-b^{\dagger}_{\downarrow}b_{\downarrow}\right)
\end{eqnarray} 
We can see right away that their expectation values are zero for the ground state, in contrast with the broken symmetry solutions of the mean field theory. We thus look up at the next moment, the spin correlation function. In particular, we can obtain the following result for the spin correlator  for the ground state
$\langle \Psi_{GS}|S_z(a) S_z(b)|\Psi_{GS}\rangle= -\frac{c_S^2}{2}$. 
Thus, for $U=0$ there is some spin correlation ($-1/8$).  As $U$ is ramped up, the correlation tends to  $-1/4$, the value  expected for $c_S=1/\sqrt{2}$ and $c_2=0$.  In that limit, wave function is identical to the spin singlet of the antiferromagnetic  Heisenberg dimer.

  In the strong coupling  limit, $\tilde U>>\tilde t$,  
  it is well known\cite{anderson59,moriya60} that the four lowest
levels in the model of equation (\ref{H2})  can be mapped into 
 the Heisenberg Hamiltonian: 
\begin{equation}
{\cal H}_{\rm Heis} = J_{H}\vec{S}_A\cdot\vec{S}_B
\label{HEIS}
\end{equation}
where $J_{H}\simeq \frac{4\tilde{t}^2}{\tilde{U}}$. The Hamiltonian of equation (\ref{HEIS})
has a ground state singlet ($S=0$) as well as an excited state triplet with $S=1$,
separated in energy by $\Delta=E(S=1)-E(S=0)=J_H$.

So, the picture that emerges from this model is the following:  both structures considered here have  two in-gap states  each. These in-gap states have a splitting $\delta$ that arises from the hybridization of 2 sublattice polarized  zero modes.  The system is thus modeled with a Hubbard dimer.  At half filling, the Hubbard dimer can be effectively mapped into a spin model, when the energy cost of double occupancy of these zero modes, given by equation (\ref{U}) is much larger than the hybridization splitting. 

The next question we address is how to up-scale these Hubbard dimers to obtain  larger structures.  In other words, how to couple more dimers together. There are at least two ways in which can do this. If we increase the width of the square shape graphene ribbon, making the zigzag edges wider,  we shall increase the number of zero modes. Eventually, this leads to the case of 1D channels with ferromagnetic interactions, discussed in the previous section.     The other way, is discussed in the next section.

 \begin{figure}[]
\begin{center}
\includegraphics[width=0.5\textwidth]{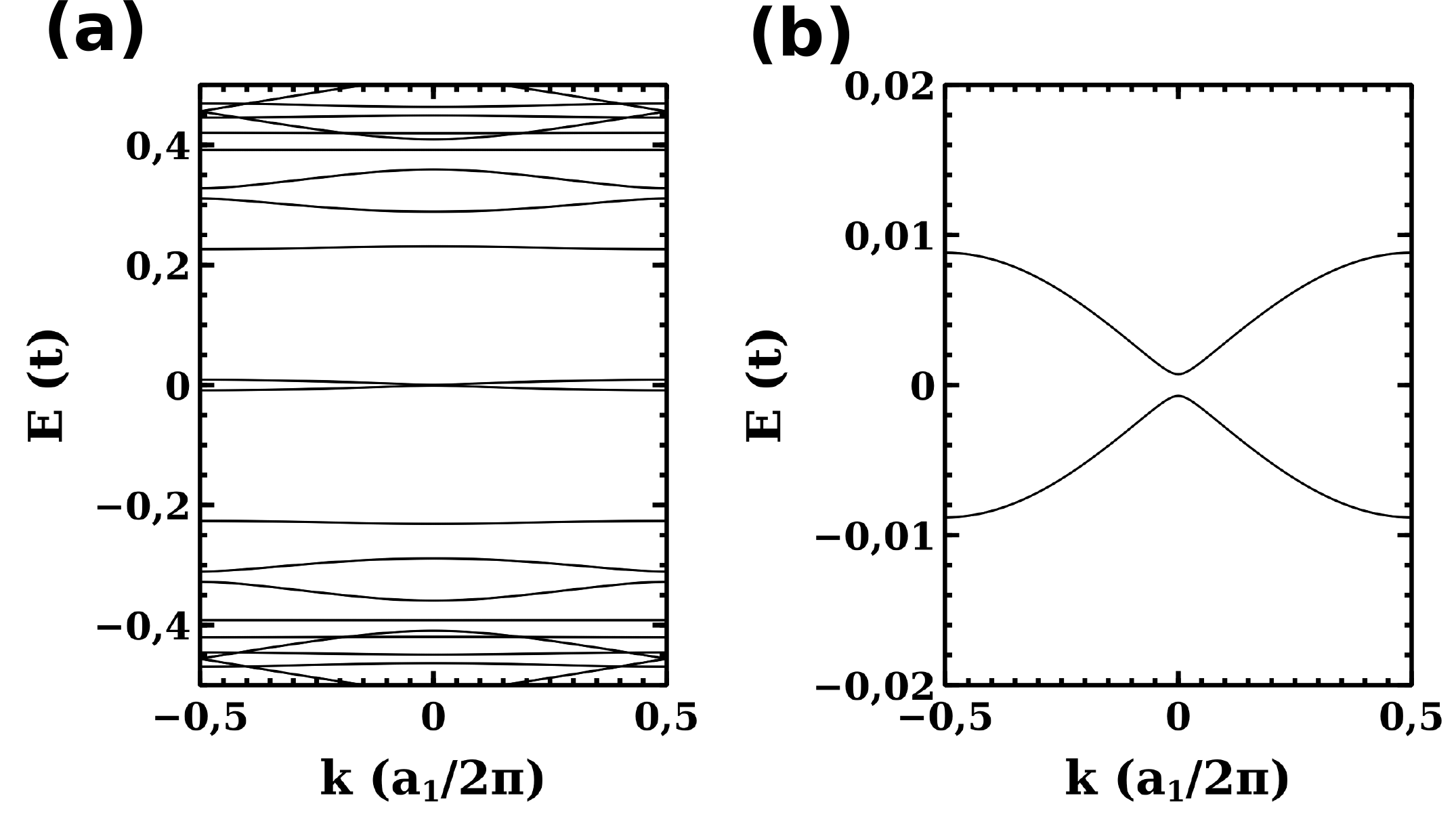}
\end{center}
	 \caption{(a) Bands of a GNR made of two armchair sections with different width  ($W_t=12$, $W_n$=8.  (b) Zoom showing the in-gap narrow bands,  that arise from the interface zero modes,  featuring  a dimerization splitting at $k=0$,  coming from the different intra and intercell hybridization of the interface zero modes.  }
         \label{bandsdimer}
\end{figure}

\section{Dimerized spin chain}
We consider   a one dimensional ribbon with armchair edges and alternating
section,  whose unit cell contains two interface (quasi)-zero modes.  The unit
cell is described by two length scales, $W_t$ and $W_n$ that describe the width
of the thicker and narrower AGNR.    These two length scales control the
effective hopping between zero modes.  The zero modes thus lead to the formation two bands, inside the gap of the AGNR, shown in figure \ref{bandsdimer}. These in-gap bands
 are effectively described by the $t,t'$ model Hamiltonian.  Their bandwidtt, governed the effective hybridization of the interface states, can
be tuned by changing $W$, and can easily much smaller than the band-gap of
these ribbons. In  figure \ref{bandsdimer}) we show a bandwidth of $0.02$t inside a gap of  0.5$t$. 

 For $t\neq t'$ the two bands have a gap at $k=\pi/L$, where $L$ is the length
 of the unit cell.  The gap closes at $t=t'$. This point separates  two
 insulating phases $t>t'$ and $t<t'$ that are topologically distinct and have a
 different Zak phase\cite{delplace11}.  As a result, only one of them has zero
 modes the edges. This happens when the last dimer is affected by the smallest
 of $t$ and $t'$

 \begin{figure*}[t!]
\begin{center}
\includegraphics[width=\textwidth]{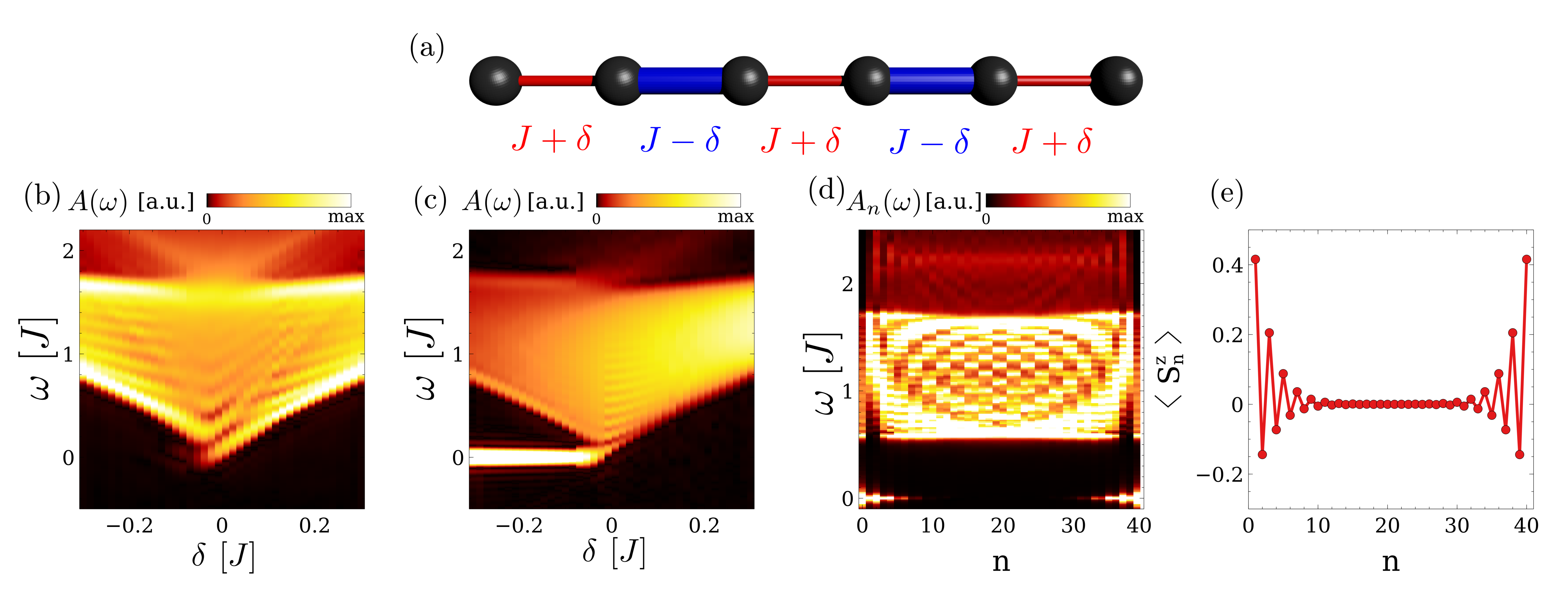}
\end{center}
	 \caption{(a) Sketch of a dimerized Heisenberg
	 $S=1/2$ spin chain, having two exchange
	 constants with different strength.
	 Bulk (b) and edge (c) spectral function of the
	 Heisenberg model, showing a gap for $\delta\ne 0$
	 in the bulk (b), while zero modes for $\delta<0$
	 on the edge.
	 Panel (d) shows the spectral function
	 in every site, showing the
	 edge localization of the zero modes. Panel (e)
	 shows the local magnetization under a small external field
	 $B_z = 0.01 J$, showing the localization
	 of the emergent edge modes.}
	 \label{figdmrg}
\end{figure*}

We now consider the effect of interactions in the strong coupling limit ${\tilde U} \gg \tilde{t},\tilde{t'}$.  Using the results of the previous section,  and in
line with previous work for other GNR structures\cite{golor14,koop17},  we
consider a spin  model\cite{chitra95}
\begin{equation}
{\cal H}_{\rm chain} =  \sum_{n=1,N}\left(J+\delta\right)  \vec{S}_{2n-1}\cdot\vec{S}_{2n}+
\left(J-\delta\right)  \vec{S}_{2n}\cdot\vec{S}_{2n+1}
\label{heischain}
\end{equation}
The different exchanges are related to the different hoppings
where we can write up:
\begin{equation}
J_n= 4\frac{\tilde{t}_n^2}{\tilde U}; \;\;
J_t= 4\frac{\tilde{t}_t^2}{\tilde U}; \;\;
\end{equation}
We now have $J_t-J_n=2\delta$ and $J_t+J_n= 2 J$.   

The previous model can be easily solved (\ref{heischain}) for
a chain of $N=40$ sites (20
dimers)  using density matrix renormalization
group\cite{PhysRevLett.69.2863,
RevModPhys.77.259,PhysRevB.66.045114,Schollwck2011,PhysRevLett.119.046401,PhysRevB.72.180403,PhysRevB.87.081106}
implemented in the
matrix product
formalism\cite{Verstraete2008,PhysRevB.87.155137,itensor} (DMRG). 
In particular, this
method
permits to obtain the expectation values and  correlation functions  of  spin
operators the ground state in a computationally efficient manner. 
The matrix product formalism also allows to access dynamical quantities
of many body systems,\cite{RevModPhys.78.275,PhysRevB.90.115124,2019arXiv190607090L}.
Specifically,  in the following we
discuss the dynamical structure factor defined as
\begin{equation}
A_n(\omega)=\langle \Psi_{GS}|S_n^z \delta\left(\omega - {\cal H}+E_{GS}\right) S_n^z|\Psi_{GS}\rangle
\end{equation}
The local dynamical structure factor can be qualitatively understood
as the quantity giving access to the local density of states of the
many body spin excitations.
Using this method, we find that, in bulk, there is a gap for $\delta\neq0$ (see
Figure \ref{figdmrg})(b).  The edge structure factor shows gapless states for
$\delta<0$
(see figure \ref{figdmrg})(c).
In figure \ref{figdmrg}(d) we show the
map of the spectral function  as a function of
position and energy, for a
 a fixed value of $\delta=-0.2 J$.  It is apparent that both edges host zero
 modes. This phenomenology is similar to the one of the SSH model, and
 therefore this system can be intuitively understood as a many-body
 version of a symmetry-protected
 topological state.\cite{PhysRevLett.42.1698}

 The edge topological excitations can be accessed by means of a
 weak external field. First, it is important to note that, for sufficiently large chains,  the dimerized Heisenberg model has a ground state that is
 four fold degenerate when $\delta<0$, consisting on a singlet
 and a triplet state, stemming from the dangling edge excitations.
 Upon an introduction of a weak external field, the triplet state
 with $S_z=+1$ becomes the ground state, giving rise to
 a finite magnetization in the edges.
 In figure \ref{figdmrg}(e) we show the local
 expectation value of $S_z$ for the $N=40$ chain
 with a small external field of $B_z = 0.01J$.

\section{Magnetic ribbons competing  a superconducting proximity effect}
In this section we discuss a situation where graphene ribbons that host local moments, driven by the exchange interactions discussed above, are in addition exposed to  superconducting proximity effect coming from the substrate.  This system permits the study of the competition between magnetism and superconductivity. 

\begin{figure}[t!]
\begin{center}
\includegraphics[width=0.49\textwidth]{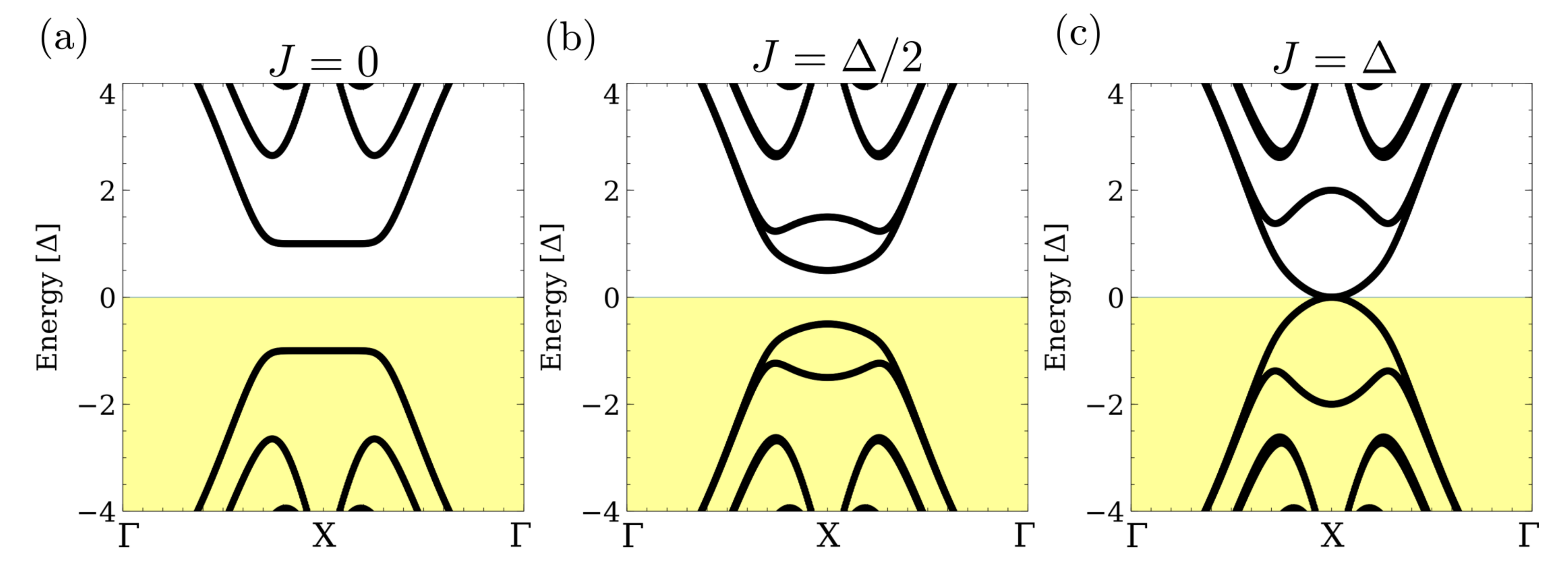}
\end{center}
	 \caption{
		 (a,b,c) Bogoliubov-de-Gennes spectra
	 for a zigzag ribbon with superconducting proximity
	 effect, for different values of the edge exchange
	 $J$, showing the closing of the
	 gap as $J$ increases. The width of the ribbon in 16 atoms,
	 the edges are in the AF configuration and we
	 took $\Delta=0.2t$}
         \label{figbandsBdG}
\end{figure}

We consider first the energy spectrum of the  ribbon without magnetic order and with a  proximity pairing $\Delta$ in the Bogoliubov -de Gennes (BdG)  Hamiltonian\cite{beenakker06,
sanjose15,lado16}.  As we show in
figure \ref{figbandsBdG}(a),  superconducting proximity  opens up a gap $\Delta$  at
the Fermi energy.  Time reversal symmetric perturbations can modify the
spectrum, but the energy levels can not be inside the gap.

\begin{figure}[t!]
\begin{center}
\includegraphics[width=0.48\textwidth]{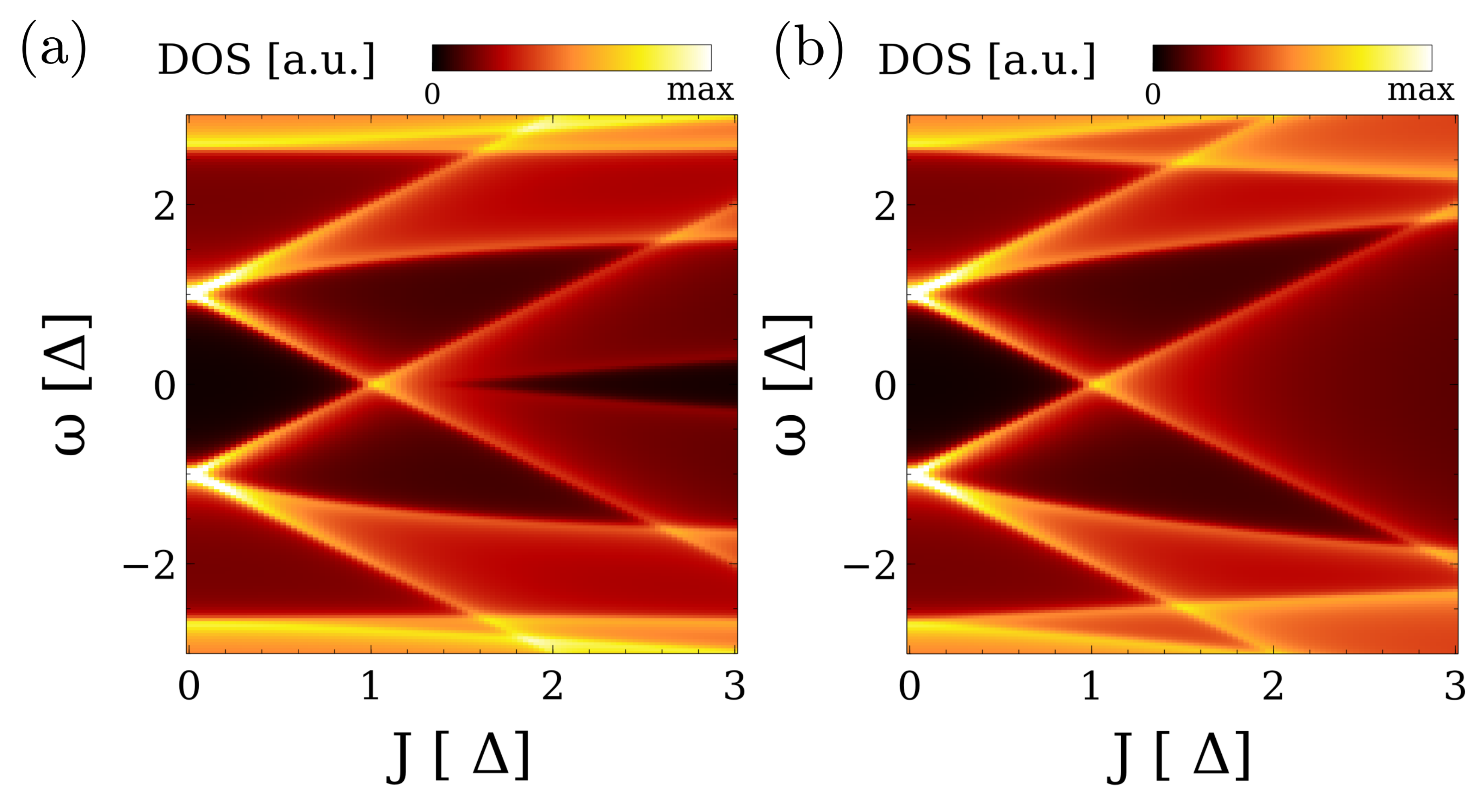}
\end{center}
\caption{Evolution of the DOS for 1D zGNR with superconducting proximity effect as the 
spin-splitting $J$ induced by ferromagnetic edge order.  Left: antiparallel inter-edge alignment. 
Right: parallel inter-edge alignment. In both instances in-gap states are induced as $J$ is ramped up. We took a width of 16 atoms
and $\Delta=0.2t$.
}
\label{fig:shibaZZ}
\end{figure}

Things are different when  we consider the effect of the magnetic order at the edges.
We follow our previous work\cite{lado16}  and we model the exchange by  adding a spin and position dependent on-site potential: 
\begin{equation}
V_{\rm exch}=  \sum_{i\in \rm{edge}} \frac{J(i)}{2} \left(n_{i\uparrow}-n_{i\downarrow}\right)
\end{equation} 
where $J(i)=0$ everywhere except at the top edge, for which  $J(i)=J$  and the bottom edge, for which  $J(i)=\pm J$.  We thus consider two different relative orientations of the edge
magnetization.  The modification of the energy bands and  density of states of
the
superconducting ribbon due to
the exchange is shown in
figures \ref{figbandsBdG} and \ref{fig:shibaZZ}, respectively.  The
most outstanding feature is the emergence of in-gap Yu-Shiba-Rusinov (YSR)
-\cite{yu65,shiba68,rusinov69} states.  Interestingly, we find an  energy
dependence of the YSR states linear in $J$, in line with the one obtained for
hydrogenated graphene\cite{lado16}, but different from the standard non-linear
dependence of YSR states in normal metals.   

The effect of the increasing exchange coupling can be also observed
in the Bogoliubov-de-Gennes excitation spectra of the ribbon as shown
in Fig. \ref{figbandsBdG}. First, in the absence of exchange field
a superconducting gap opens up. As the exchange field increases, the
gap starts closing until at critical value the system becomes gapless.
This phenomenology is similar to the one found in single
magnetic impurities,\cite{RevModPhys.78.373}
where as $J$ is increased a single
in-gap excitation approaches the charge neutrality point,
giving rise to a parity switching point. In the case of graphene
nanoribbons the behavior is more complex due to
the existence of several YSR branches that give rise a to a continuum
of in-gap excitations.\cite{NadjPerge2014,rontynen15,Mnard2015,
PhysRevLett.117.186801,Kezilebieke2018}

\section{Experimental probes}
The experimental study of the non-trivial electronic phases discussed in this
chapter can be carried out with state of the art scanning tunnelling microscope (STM) spectroscopy.  More
specifically,  the collective  spin excitations of  either the magnetically
ordered phase, or those of the spin-liquid phase of the dimerized spin chain
could be probed with inelastic electron tunneling spectroscopy (IETS)\cite{hirjibehedin06,fernandez09,li19}.  Spin
excitations in the range of a few meV can be resolved, and the mapping of their
intensity  profile across the ribbons can help to discriminate from other
inelastic excitations in the system, such as phonons.

Given that both the synthesis and the STM probing require to have the GNR on
top of a conducting surface,  there will be spin exchange interactions between
the
local moments at graphene and the conduction electrons at the surface, that we have ignored so far.  
These
Kondo interactions can compete with the exchange interactions discussed so far
in several ways. First, if sufficiently strong, Kondo effect could screen the
local moments in graphene.  Recent experiments in GNR studied with STM
spectroscopy show Kondo peaks\cite{li19}, that very likely imply the screening
of the graphene local moment. Second, additional indirect exchange
interactions, mediated by the substrate electrons, can compete, or perhaps
enhance, the graphene mediated interactions.   In any event, these Kondo
interactions will affect the lineshapes of the inelastic electron tunneling
spectra measured with STM \cite{zhang13}

\section{Conclusions and outlook}
We have explored the several examples of  emergence of non-trivial quantum phases in graphene nanoribbons. 
The building blocks for these phases are zero modes that form narrow bands at the Dirac energy. The band-width of these bands  depends on the aspect ratio of the structures, which provides thereby a control knob.  We have focused mostly on the case of neutral GNR, that leads to half-full narrow bands  that result in insulating structure with interesting spin physics but frozen charge dynamics.    Departure from half-filling is expected to result in very interesting electronic properties. For instance, doping the 1D ferromagnetic edge is expected to result in domain walls that host fractionalized electrons\cite{lopez17}.  Doping the dimerized spin chains  might result in superconducting phases. 
Superconducting proximity effect can be another way to explore the interplay between spin and charge.   We have discussed the emergence of in-gap Shiba bands in graphene ribbons.
When  non-collinear magnetic ground states and/or spin-orbit coupling are considered,  the Shiba bands could  give rise to topological superconductivity with Majorana end modes  \cite{NadjPerge2014}.

{\em Acknowledgments}
 J. F.-R. acknowledge financial support
from FCT for the P2020-PTDC/FIS-NAN/4662/2014,
 the P2020-PTDC/FIS-NAN/3668/2014 and the UTAPEXPL/NTec/0046/2017 projects, as well as Generalitat
Valenciana funding Prometeo2017/139 and MINECO Spain (Grant No. MAT2016-78625-C2).
 R.O.C. acknowledges ``Generalitat 
Valenciana'' and ``Fondo Social Europeo'' for a Ph.D. fellowship (ACIF/2018/198).
J. L. L. acknowledges financial support from the ETH Fellowship program.

\bibliographystyle{apsrev4-1}
\bibliography{References}{}
\end{document}